\documentclass{article}
\usepackage{psfig}
\usepackage{astro}
\usepackage{amsmath}

\newcommand\degd{\ifmmode^{\circ}\!\!\!.\,\else$^{\circ}\!\!\!.\,$\fi}

\begin{document}
\begin{center}{\LARGE{\bf Polarization in Sagittarius A*}\\
 Geoffrey Bower}\end{center}

\begin{center} {\it Invited Article, GCNEWS, Vol 11, p. 4 (2000)}\end{center}

\begin{center}{\bf Abstract}\end{center}
\begin{quotation}

We summarize the current state of polarization observations of Sagittarius A*,
the compact radio source and supermassive black hole candidate in the
Galactic Center.  These observations are providing new tools for
understanding accretion disks, jets and their environments.
Linear polarization observations have shown that
Sgr A* is unpolarized at frequencies as high as 86 GHz.  However, recent
single-dish observations indicate that Sgr A* may have strong linear
polarization at frequencies higher than 150 GHz.  Circular polarization,
which was recently discovered in Sgr A*, is detectable to 
frequencies as high as 43 GHz.  It is strongly variable on a 10-day
timescale, with the degree of variability increasing with frequency.
At low frequencies, fractional circular polarization is stable over
nearly 20 years.  We discuss some of the possible models for
the origin of the circular polarization.

\end{quotation}

\section*{Introduction}
Polarization observations have been very successful in the study of
active galactic nuclei (AGN).  Linear polarization (LP) observations
have demonstrated conclusively that synchrotron emission is the
dominant mechanism in jets, that magnetic fields are present in jets,
and that shocks propagate in jets leading to variability in the total
and polarized intensity.  Circular polarization (CP) observations,
while producing less clear cut conclusions so far, have the potential
for illuminating new physics.  For example, on the basis of CP
observations Wardle \etal\ (1998) have recently claimed that jets are
not primarily hadronic but leptonic.

With the success of polarization observations in AGN in mind, we set
out to determine the polarization characteristics of Sagittarius A*,
the supermassive black hole candidate and compact radio source in the
Galactic Center.  Polarization observations offer the possibility to
break the theoretical deadlock that exists over the nature of Sgr A*.
Both advection dominated accretion flow (ADAF) models (e.g., \"{O}zel,
Psaltis \& Narayan 2000) and accretion-disk-powered-jet models (Falcke
\& Biermann 1999) can account for the spectrum from centimeter
wavelengths to the X-ray, while meeting the size constraints imposed
by very long baseline interferometry.

Observational methods are limited by the high galactic extinction and
the hyperstrong interstellar scattering screen in the Galactic Center.
Polarization observations have the potential to see through the
scattering screen and determine intrinsic source characteristics.
Now, new observational results are leading to new opportunities for
understanding of Sgr A*.

\section*{Linear Polarization}

Sgr A* has been known to show no LP since not long after its discovery
(Balick \& Brown 1974).  However, this result had not been quantified
and it was often assumed that the interstellar scattering screen
depolarized the intrinsic source through beam or bandwidth
depolarization.  Recently, with VLA and BIMA observations, we
confirmed the low LP and showed that interstellar depolarization is
unimportant (Bower \etal\ 1999ab).  We summarize in Figure~1 our
continuum measurements of LP in Sgr A*.  Note that the increasing
value of the upper limits with frequency is a measure of the
increasing difficulty of these observations.

Both bandwidth and beam depolarization are Faraday effects.  The
dense, magnetized plasma of the interstellar scattering screen is a
potential source of these effects.  Observed rotation measures (RMs)
in the GC are on the order of $10^3 \rdm$ (Yusef-Zadeh, Wardle \&
Parastaran 1997) but they could potentially reach $10^4 - 10^5\rdm$
(Bower \etal\ 1999a).  High frequency polarimetry and
spectro-polarimetry are useful techniques for avoiding these effects.
The first relies on the decreasing strength of the effects with
frequency.  The second removes the effect of large RMs through a
Fourier transform method.

In the case of bandwidth depolarization, a large RM rotates the LP
position angle through $\ga \pi {\rm\ rad}$ in the observing band.
High frequency continuum polarimetry and spectro-polarimetry rule this
effect out in Sgr A*.  The RM must exceed $8\times 10^6\rdm$ in order
to depolarize any intrinsic LP at 43 GHz in the 50 MHz bandwidth of
the VLA.  VLA spectro-polarimetry at 4.8 and 8.4 GHz also place a
lower limit of $\sim 10^7\rdm$ to an RM that depolarizes intrinsic LP.

Beam depolarization occurs when different paths of the scatter
broadened image experience different RMs.  This places stricter limits
on the RM than bandwidth depolarization, if the scale of turbulence is
sufficiently small.  RM variations greater than $3\times 10^5\rdm$ are
necessary to depolarize the signal at 86 GHz.  These are disallowed,
however, by the known conditions of the scattering medium.

Very recently, Aitken \etal\ (2000) reported that with SCUBA
observations at the JCMT they have detected LP in Sgr A* at
frequencies greater than 150 GHz.  These results are complicated by
the large primary beam of the JCMT ($34^{\prime\prime}$ at 150 GHz)
and the presence of strong free-free emission and polarized dust
emission.  Removing these effects, Aitken \etal\ find $10^{+9}_{-4}\%$
LP at 150 GHz.  This result requires confirmation with a
millimeter/submillimeter interferometer.  If true, they are a
significant and challenging addition to our understanding of the
polarization properties of Sgr A*.

The very sharp rise in LP from 86 to 150 GHz may be due to Faraday
effects in the accretion region (Bower \etal\ 1999a).  Quataert and
Gruzinov (2000) have demonstrated that a high frequency detection of
LP can distinguish between disparate models for Sgr A*.  ADAF inflows
have high electron and magnetic field densities leading to bandwidth
depolarization at radii $r\la 10^4 r_g$ even at frequencies greater
than 100 GHz.  The high frequency detection by Aitken \etal\ implies
that the electron density and accretion rate must be much lower than
in the standard ADAF case.

\begin{figure}[ht]
\psfig{figure=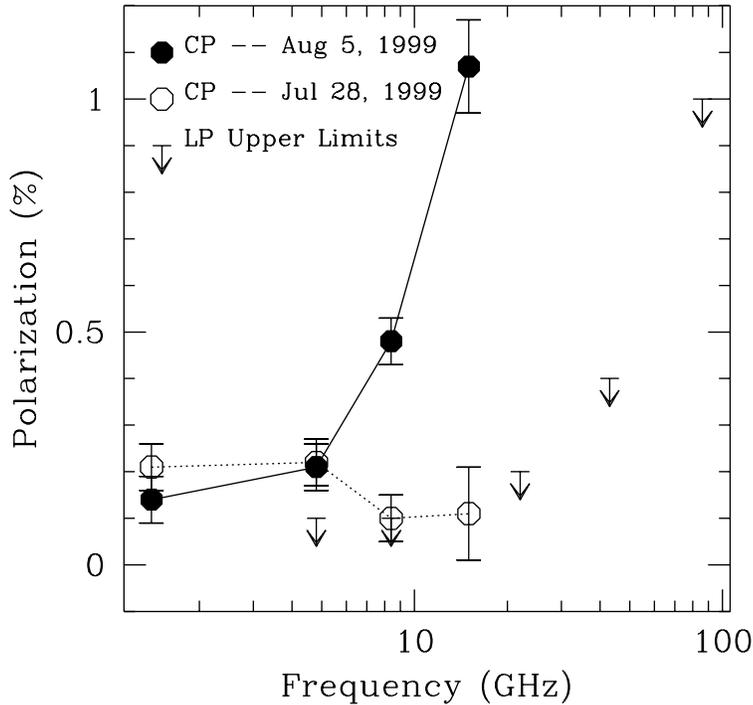,height=4in}
\caption{Linear and circular polarization in Sgr A*
from 1.4 to 86 GHz. The down arrows indicate upper limits for
linear polarization measurements.  The open octagons are CP
measurements from the VLA on July 28, 1999.  The filled octagons are
measurements from the VLA on August 5, 1999.
The sign of CP has been flipped in this figure.
Not shown is the $10^{+9}_{-4}\%$ detection of LP at 150 GHz
by Aitken \etal}
\label{fig2}
\end{figure}

\section*{Circular Polarization}

In the course of discovering that Sgr A* shows no LP at centimeter
wavelengths, we discovered serendipitously that Sgr A* has strong CP
(Bower, Falcke \& Backer 1999c).  This is surprising in light of the
strong limits on linear polarization.  AGN typically have LP-to-CP
ratios $\gg 1$ (Weiler \& de Pater 1983).  But for Sgr A* at 4.8 GHz,
the fractional CP is $\sim -0.3\%$, which is at least 3 times the
level of LP.  We have since followed up with monitoring campaigns
using the VLA and the ATCA and an analysis of data from the VLA
archives covering 18 years.  CP measurements with the VLA are technically
difficult; confirmation with ATCA observations (Sault \& Macquart
1999) and at multiple frequencies over many epochs with the VLA has
been crucial.

The VLA archive data clearly indicates that the CP flux at 4.8 GHz is
stable at $-0.31\pm 0.13\%$ over 18 years, despite a factor of two
change in total intensity.  CP at 8.4 GHz is also apparently stable at
$-0.27\pm 0.10\%$ over 10 years.  However, short-term variability around
the mean value is apparent.  In fact, the CP flux changes
significantly on timescales as short as a few days.  In Figure~1, we
show the spectrum of CP from 1.4 to 15 GHz on two dates separated by a
week.  In this short interval, the CP has increased dramatically at
frequencies greater than 5 GHz.  This flare in fractional CP is
coincident with a 30\% flare in total intensity (Bower, Falcke, Sault
\& Backer 2000).

In general, the degree of CP variability increases sharply with
frequency.  Recently, after several epochs without detection, we have
detected strong CP with an inverted spectrum at 22 and 43 GHz using
the VLA.  The analysis of this new result is not yet complete but the
preliminary result suggests that the CP spectrum rises during flares
into the millimeter regime.

Interpretation of the CP results is not yet certain.  The variability
suggests a two component model that includes a steep-spectrum
quiescent phase and frequent inverted-spectrum flares.  Under this
scenario there are two distinct origins for the CP, originating in
the low and high frequency components of Sgr A*.  In jet models, these
are more and less compact regions, respectively.  In ADAF models, the
high/low frequency component originates from
non-relativistic/relativistic electrons in the flow.

Gyrosynchrotron radiation is a natural model for the high frequency
component of the CP.  It originates from non-relativistic electrons in
a strong field and produces low LP-to-CP ratios.  However, this is
certainly not the only possible model.  Synchrotron radiation can
produce low LP-to-CP ratios under certain conditions.  This can
include the effects of linear-to-circular conversion, which is
believed to be active in 3C 279 (Wardle \etal\ 1998).  Finally,
birefringent scattering, in which a magnetized region scatters LCP and
RCP differentially in angle, could play role if the natural steep
spectrum that it produces can be accounted for.  Any model must
explain the low LP-to-CP ratio, the inverted spectrum of CP and the
nature of the variability.  Detailed modeling of the source is
probably necessary.

Outstanding observational issues for the CP in Sgr A* are the
relationship between CP and total intensity fluctuations; determining
the turnover frequency of CP flares; and observing potential
structural changes on the AU scale that are related to CP variability.

In the future, CP and LP measurements could be very important tools
for understanding accretion flows, jets and their environments.
Nearby low luminosity AGN, which are often associated with ADAF
systems, are the first targets for these studies.  However, before
that we must ground our understanding of these behaviors in the best
studied and most well-known AGN, Sgr A*.

\section*{References}

\parindent 0in

Aitken, D.K., Greaves, J.S., Chyrsostomou, A., Holland, W.S.,
Hough, J.H., Pierce-Price, D. \& Richer, J.S.,  2000, \apjl, accepted

Balick, B. \& Brown, R.L., 1974, \apj, 194, 265

Bower, G.C., Backer, D.C.,  Zhao, J.-H., Goss, M. \& Falcke, H., 1999a,
\apj, 521, 582

Bower, G.C., Wright, M.C.H., Backer, D.C. \& Falcke, H., 1999b,
\apj, 527, 851

Bower, G.C., Falcke, H. \& Backer, D.C., 1999c, \apjl, 523, L29

Bower, G.C., Falcke, H., Sault, R. \& Backer, D.C., 2000, in preparation

Falcke, H. \& Biermann, P.L., 1999, \aap, 342, 49

\"{O}zel, F., Psaltis, D. \& Narayan, R., 2000 \apj, accepted

Quataert, E. \& Gruzinov, A., 2000, \apjl, submitted

Sault, R.J. \& Macquart, J.-P., 1999, \apjl, 526, L85

Wardle, J.F.C., Homan, D.C., Ojha, R., \& Roberts, D.H., 1998, Nature,
395, 457

Weiler, K.W. \& de Pater, I., 1983, \apjs, 52, 293

Yusef-Zadeh, F., Wardle, M. \& Parastaran, P., 1997, \apjl, 475, L119                            
    
\end{document}